# Semiconductor Spintronics:
# Progress and Challenges

**Emmanuel I. Rashba**
*Department of Physics, Harvard University, Cambridge,
MA 02138, U.S.A.*

## 1. Introduction

Spin is the only internal degree of freedom of electron, and utilizing it in the new generations of semiconductor devices is the main goal of semiconductor spintronics. Contemporary semiconductor electronics is based on electron charge only. It is expected that involving electron spin will provide electronic devices with new functionalities, and achieving quantum computing with electron spins is among the most ambitious goals of spintronics.[1,2] During the last five years, there has been impressing progress in this field, both in experiment and in developing theoretical concepts. Because goals are highly challenging, and numerous difficult problems should be solved, research is developing along several avenues. E.g., increase is the spin coherence time of gate-controlled double quantum dots by several orders of magnitude has been achieved recently by applying spin echo technique;[3] it is promising for the prospects of spin computing with quantum dots. In what follows, we concentrate on fundamentals and recent developments related to a different branch of spintronics that is concerned with employing spin-orbit coupling for achieving direct electrical control of electron spins in semiconductor nanostructures. As compared with magnetic control of spins, electric control has prospects of being more efficient and allowing access to electron spins at nanoscale.

Strong enhancement of spin-orbit coupling in crystals as compared to vacuum originates from the effect of large gradients $\nabla V(r)$ of the crystal field and high electron velocities $v$ near nuclei, and is imprinted in the wave functions and energy spectrum of Bloch states. In vacuum, the dimensionless parameter of spin-orbit coupling is about $E(k)/m_0 c^2 \sim 10^{-6}$, with the electron energy $E(k) \sim 1$ eV and the Dirac gap $m_0 c^2 \sim 1$ MeV. In semiconductors, the similar parameter is $\Delta_{SO}/E_G \sim 1$, $\Delta_{SO} \sim 1$ eV being the spin-orbit splitting of valence bands and the forbidden gap $E_G \sim 1$ eV. This enhancement makes semiconductors promising media for electrical manipulating electron spins.[4]

## 2. Basic concepts of semiconductor spintronics

Apparently, the first practical application of electrically-driven spin transitions belongs to laser physics and is dated as early as 1971.[5] More recent research was initiated by the 1990 paper by Datta and Das who advanced the idea of a spin-transport device based on spin interference in media with spin-orbit coupling.[6] Afterwards, it became known as spin transistor. Despite the fact that the attempts of creating such a device have not been successful by now, and its feasibility has been questioned,[7] the basic principles underlying it strongly influenced following research.

A toy-model spin-orbit Hamiltonian describing electrons in asymmetric two-dimensional (2D) systems (Rashba term[8,9]) is

$$H_\alpha = \alpha(\boldsymbol{\sigma} \times \boldsymbol{k}) \cdot \mathbf{z}_0, \tag{1}$$

where $\alpha$ is a spin-orbit coupling constant, $\boldsymbol{\sigma}$ is a Pauli matrices vector, $\boldsymbol{k}$ is electron wave vector, and $\mathbf{z}_0$ is a unit vector perpendicular to the confinement plane. When rewritten as

$$H_\alpha = g\mu_B(\boldsymbol{B}_\alpha \times \boldsymbol{\sigma})/2, \quad \boldsymbol{B}_\alpha(\boldsymbol{k}) = (2\alpha/g\mu_B)(\boldsymbol{k} \times \mathbf{z}_0), \tag{2}$$

where $\mu_B$ is the Bohr magneton and $\boldsymbol{B}_\alpha(\boldsymbol{k})$ is an effective momentum-dependent spin-orbit field, the Hamiltonian $H_\alpha$ describes spin precession in the field $\boldsymbol{B}_\alpha(\boldsymbol{k})$. The same phenomenon can be also understood in terms of two eigenstates of the Hamiltonian $H_\alpha$ with the same propagation direction $\mathbf{k}_0$ and energy $\varepsilon$, but with different momenta $k_\pm$ depending on the spin-orbit coupling constant $\alpha$ (spin birefringence). Therefore, if spin-polarized electrons are injected at $x = 0$ along the direction $\mathbf{k}_0$ in a spin state that is not an eigenstate for the field $\boldsymbol{B}_\alpha(\boldsymbol{k})$, the resistance of the device is controlled by the $\alpha$-dependent phase of the electron wave function near the spin-polarized drain at $x = L$.

The Datta and Das device is based on the following principles:

(a) Spin injection from a ferromagnetic source and spin detection by a ferromagnetic drain,
(b) Electrical control of spin-orbit coupling $\alpha$ by a Schottky gate,
(c) Spin precession in the spin-orbit field $\boldsymbol{B}_\alpha$, and
(d) Spin interference.

Lately, there was impressing progress in developing ferromagnetic injectors, including better understanding of the role of contacts between spin injectors and semiconductor microstructures. Meantime, the paradigm shifted, and a lot of attention has been paid to generating and injecting spin populations all-electrically, by means of spin-orbit coupling. Avoiding ferromagnetic elements would allow eliminating stray magnetic fields. Electrical control of spin-orbit coupling[10,11] and spin precession[12] in the field $\boldsymbol{B}_\alpha$ have been reported long ago, while spin interference has been observed only recently, see Sec. 4 below.

Developing all-semiconductor electrically controlled spintronics needs better understanding spin transport in media with spin-orbit coupling that is rather nontrivial. In what follows, we review some of the recent progress in this field.

### 3. Where spin coupling to external electric field comes from?

Electrically induced quantum transitions are usually described in terms of oscillator strengths that are subject to the oscillator sum rule (Thomas-Kuhn-Reiche theorem). It follows from the standard commutation relation

$$i[k, x] = 1. \tag{3}$$

When the commutator is written as a sum over the intermediate states, it becomes a sum of the terms

$$f_{n \leftarrow \ell} = i\{<\ell| k |n><n| x |\ell> - <\ell| x |n><n| k |\ell>\}, \tag{4}$$

which are oscillator strengths of $n \leftarrow \ell$ transitions. In the absence of spin-orbit coupling, calculating the commutator of the coordinate $x$ and the Hamiltonian, we find

$$<n| x |\ell> (E_\ell - E_n) = i\hbar^2 <n| k |\ell> / m_0, \tag{5}$$

and after substituting it into Eq. (4) we arrive at

$$f_{n \leftarrow \ell} = (2\hbar^2/m_0) |<\ell| k |n>|^2 / (E_n - E_\ell), \tag{6}$$

where $m_0$ is the electron mass in vacuum. For local states, the oscillator sum rule

$$\sum_n f_{n \leftarrow \ell} = 1 \tag{7}$$

includes nondiagonal terms, $n \neq \ell$, only. Indeed, all diagonal terms $n = \ell$ vanish because matrix elements of the coordinate $x$ in Eq. (5) are finite.

However, because Bloch states are extended, diagonal matrix elements of $x$ diverge. Hence, diagonal matrix elements of $k$ may survive. If one takes into account that the oscillator strengths $f_{n \leftarrow \ell}$ of Eq. (6) coincide, with the accuracy to a factor $m_0$, with the summands in the standard expression of $k \cdot p$ theory for the inverse effective mass $m_\ell$ in the $\ell$-th Bloch band, one arrives at the equation

$$m_0/m_\ell + \sum_{n \neq \ell} f_{n \leftarrow \ell} = 1. \tag{8}$$

Therefore, $m_0/m_\ell$ is the oscillator strength $f_{\ell \leftarrow \ell}$ for the transition from the state $\ell$ "into itself".[13] It is exactly the oscillator strength that manifests itself in the Drude and cyclotron absorption.

The problem is what happens to this oscillator strength in a noncentrosymmetric system when spin-orbit coupling enters into the game, $\alpha \neq 0$, and a spin degenerate band splits into two subbands. This situation is shown in Fig. 1. For each state, the total oscillator strength $m_0/m_\ell$ is divided between the transition "into itself" and the transition between branches. For the transitions from the spectrum bottom, the inter-branch transition energy equals $2E_{so}$, with $E_{so} = m\alpha^2/\hbar^2$, and the oscillator strength is divided equally between both transitions. For a given wave vector $\boldsymbol{k}$, electron spins have opposite directions on two spectrum branches, hence, inter-branch transitions are spin-flip transitions. Meantime, they have tremendous intensities that are comparable to the intensity of the cyclotron resonance. With increasing electron energy, intensities of inter-branch transitions decrease, but only as $k_\alpha / k_F$, where $k_\alpha = m\alpha/\hbar^2$ is the spin precession momentum in the field $\boldsymbol{B}_\alpha(\boldsymbol{k})$, and $k_F$ is Fermi momentum. Therefore, their intensities remain high for reasonable $\alpha$ values. Outside the spectral region of interbranch transitions, their Kramers-Kronig transform describes spin coupling to electric fields;[14] spectral dependence of the corresponding responses can be only found from detailed transport equations. In a strong magnetic field $\boldsymbol{B}$, inter-branch transitions transform into the Electric Dipole Spin Resonance (EDSR), whose intensity is usually much higher than the intensity of Electron Paramagnetic Resonance (EPR).[4,8]

The important role that intrabranch transitions play in spin transport will be discussed in Sec. 6 below.

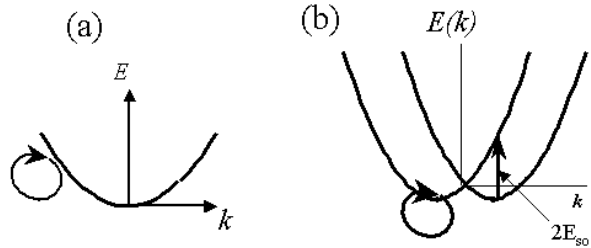

**Figure 1.** Spin-degenerate spectrum in the absence of spin orbit coupling (a) and spin split spectrum of 2D electrons with spin-orbit coupling of Eq. (1) (b). In (a), the oscillator strength for transitions "into itself" (a circle with an arrow) equals $m_0/m_\ell$. In (b), this oscillator strength is divided between transitions "into itself" (a circle with an arrow) and transitions between two branches (a vertical arrow). For the transitions from the spectrum bottom (shown in the figure), the oscillator strength $m_0/m_\ell$ is equally divided between both transitions.

## 4. Experimental achievements: Spin populations and spin interference

Because of the spin coupling to electric field, propagation of electric current across a sample is accompanied by spin accumulation in the bulk of three-dimensional (3D) and 2D systems.[15,16] For thin 3D layers and 2D systems, it was recently observed by Kato *et al.*,[17] Silov *et al.*,[18] and Ganichev *et al.*[19] Another related phenomenon is spin Hall effect[20-22] that manifests itself in spin accumulation near the flanks of a sample, for review see Refs. 23 and 24. Spin Hall effect observed in *n*-GaAs[25] was attributed to the extrinsic mechanism, while the effect observed in *p*-GaAs[26] to the intrinsic one. This terminology implies that the first one originates from impurity scattering and the second one from spin-orbit coupling in the bulk.

Spin interference phenomena, besides their promise for applications, are important from the fundamental point of view because they are related to quantum phases that essentially depend on the shape of the electron paths, in particular, they differ for rings where electron motion is close to adiabatic and polygons where motion near vertices is strongly nonadiabatic. Spin interference on a large array of InGaAs square loops was reported by Koga *et al.*,[27] and on a single HgTe/HgCdTe ring by Koenig *et al.*[28]

## 5. Enhancing spin responses to electric fields

Long delay in the experimental observation of electrically driven spin populations in 2D systems after their prediction was caused by a small magnitude of these effects hindering their application for semiconductor devices. In this Section, we discuss some options for enhancing spin responses.

(a) Spin response to an inhomogeneous dc field $\boldsymbol{E}(\boldsymbol{r}) = \boldsymbol{E}\exp(i\boldsymbol{q}\cdot\boldsymbol{r})$ diverges when $q \to 2k_\alpha$.[29] This behavior can be easily understood if one takes into account that adding wave vector $\boldsymbol{q}$ results in a mutual displacement of two Fermi surfaces with the same energy, and for $q = 2k_\alpha$ these surfaces touch. This results in "*spin breakdown*" because spin can be flipped with no energy price. Therefore, one should choose inhomogeneous fields with large $q \approx 2k_\alpha$ spectral components. It also suggests that optimal sizes of elements with $\alpha \neq 0$ (rings or diamonds) designed for injecting spins into $\alpha = 0$ wires should be about the spin precession length $\ell_\alpha = 1/k_\alpha$.

(b) Frequencies of time dependent fields should be close to resonances, i.e., for $B = 0$ the frequency should be about $2\alpha k_F$, and under the EDSR conditions it should be close to the EDSR frequency $g\mu_B B$. It is also important that strong scattering, $\hbar/\tau \gg \alpha k_F$, $\tau$ being the momentum relaxation time, results in the narrowing of EDSR line to the inverse spin relaxation time, $\tau_s^{-1}$.[30] The explicit shape of the line, with $\tau_s$ equal to the Dyakonov-Perel relaxation time,[31] has been recently derived by Duckheim and Loss.[32]

(c) In quantum wells, the in-plane polarization of the field $\boldsymbol{E}$ is more efficient than the out-of-plane polarization by a factor $(\omega_0^2/\omega_c\omega_s)^2$, where $\omega_0$, $\omega_c$, and $\omega_s$ are, respectively, the confinement frequency and the frequencies of the cyclotron and spin resonances.[33]

(d) Using *p*-type materials.[34]

(e) Surface states on the (111) face of Bi,[35] and on Bi/Ag(111) monolayer alloys[36] show giant spin-orbit splittings with $E_{so}$ up to 0.4 eV. Thin layers of such materials can provide ultra-short spin precession lengths $\ell_\alpha$.

New options arise if two different spin-orbit coupling mechanisms are combined. The symmetry group $C_{2v}$ of (001) quantum wells in $A_3B_5$ materials, in addition to the invariant of Eq. (1), has a different linear in $k$ invariant (Dresselhaus term)

$$H_\beta = \beta(\sigma_x k_x - \sigma_y k_y). \tag{9}$$

Pikus noticed that the 3D Dresselhaus $k^3$ - spin splitting[37] reduces to Eq. (9) in the limit of narrow quantum wells.[38] Combining $H_\alpha$ and $H_\beta$ provides new options for spintronic devices,[39,40] especially in the vicinity of the magic points $\alpha = \pm \beta$. In these points, stable spin superstructures with a $k_\alpha$ dependent period have been predicted recently.[41]

## 6. Conceptual theoretical problems

Generation of spin populations by a driving electric field is possible only due to spin nonconservation. As a result, theory of spin transport essentially differs from theory of charge transport. This difference is already obvious from Maxwellian equations that include four electric variables *E*, *D*, charge density $\rho$ and current *J*, but only two magnetic variables, *B* and *H* [or magnetization *M* = (*B* − *H*)/$4\pi$]. Therefore, absence of magnetic monopoles results not only in the absence of a magnetic analog of $\rho$, but also in the absence of magnetization current. Introduction of such a current is justified only under some special conditions, particularly, in the framework of the Mott two-fluid theory of electron transport in ferromagnets *without* spin-orbit coupling.[42] Spin-orbit coupling results in spin nonconservation. As a result, time derivative of spin magnetization, $\partial S/\partial t$, cannot be represented as a divergence of any vector. Therefore, there is no unambiguous definition of spin current, and the form of the extra term depends on the spin current definition; this term is known as torque.[43] Usually, spin current $j_i^\ell$ is defined as

$$j_i^\ell = \tfrac{1}{2} <v_i \sigma_\ell + \sigma_\ell v_i>, \tag{10}$$

where an anticommutator is taken because in the media with spin orbit coupling the velocity *v* depends on Pauli matrices $\sigma_\ell$, and $< \ldots >$ stands for averaging over the electron distribution; however, different definitions for $j_i^\ell$ have also been proposed.[44] The notion of spin currents has been used in literature for long, but it attracted more attention after Murakami *et al*.[45] and Sinova *et al*.[46] reported some unexpected properties of these currents for 3D holes and 2D electrons, respectively. Since then, these quantities became a popular playground for comparing spin responses of particles described by various spin-orbit Hamiltonians to dc and ac electric fields.

In particular, dc spin-Hall conductivity defined as $\sigma_{SH} = j_x^z/E_y$, when calculated by Kubo formula for a perfect system with spin-orbit Hamiltonian $H_\alpha$, equals $\sigma_{SH} = e/4\pi\hbar$ for arbitrary chemical potential $\mu > 0$. This "universal conductivity" raised hopes that there might be possible to find some simple results for spin accumulation near the sample flanks; spin accumulations are the only quantities currently accessible for experimental detection. However, calculation of $\sigma_{SH}$ with a proper account of electron scattering has shown that $\sigma_{SH}$ vanishes, for review see Refs. 23 and 24. The simplest formal argument, explaining this spin current cancellation, was provided by Dimitrova[47] who noticed that $j_x^z$ is proportional to the mean value of the derivate $d\sigma_y/dt$ that should vanish in a stationary state. A different argument demonstrating this cancellation is also related to the form of the free Hamiltonian only, irrespective to the potentials of non-magnetic scatterers, and is based on vanishing the spin current $j_x^z$ in a perfect sample subject to an external magnetic field perpendicular to the confinement plane.[48] Physically, vanishing of $\sigma_{SH}$ comes from the fact that there exists an intrabranch contribution to $\sigma_{SH}$, similar to the intrabranch oscillator strength of Sec. 3, that cancels the universal contribution $e/4\pi\hbar$. From this standpoint, impurities and magnetic field play a similar role: by violating momentum conservation, they unveil the intrabranch contribution.

It is currently well understood that the above cancellation is an exceptional property of the terms $H_\alpha$ and $H_\beta$ in conjunction with a quadratic nonrelativistic Hamiltonian $H_0 = \hbar^2 k^2/2m$, and it underscores the fact that while spin responses to electric fields *per se* originate from spin-orbit coupling built-in in the free electron Hamiltonian (Sec. 3), their specific form can be found only by rigorous solving proper transport equations.

Boltzmann equations for systems with spin-split energy spectrum were derived in a number of papers.[49-51] In principle, they allow solving transport problems for arbitrary value of the parameter $\alpha k_F\tau/\hbar$, but they were usually solved in the diffusive limit $\alpha k_F\tau/\hbar \ll 1$.[43,52,53] In this limit, the problem of boundary conditions becomes nontrivial because of spin nonconservation. Indeed, spin is not conserved even on a perfect boundary of $\alpha \neq 0$ and $\alpha = 0$ regions because the currents defined by Eq. (10) persist in thermodynamic equilibrium in the $\alpha \neq 0$ region but vanish in the adjacent $\alpha = 0$ region.[54] Numerical work shows that these "equilibrium currents," that are not related to any real spin transport, are especially strong near boundaries.[55] Therefore, boundary conditions for diffusive equations cannot be found from spin conservation conditions but only from consistent solving transport equations near boundaries at the scale small compared with the spin diffusion length $L_{sd}$ that for the Dyakonov-Perel spin relaxation mechanism[31] is about $\ell_\alpha$, $L_{sd} \approx \ell_\alpha$. This problem is still waiting its solution.[56] For an $H_\alpha$ semiconductor, it is expected that a dc current flowing along a perfect hard-wall boundary would produce only tiny spin accumulation near the edge.[52,57]

A different problem concerns with the relative role of extrinsic and intrinsic mechanisms and their interplay. Extrinsic mechanisms are related to the impurity scattering and are traditionally discussed in terms of skew scattering and side jump contributions. Intrinsic mechanisms are usually attributed to the spin orbit coupling terms in the Hamiltonian $H(\mathbf{k})$. A similar problem has existed in the theory of Anomalous Hall Effect (AHE) that has already a more than 50-year long history,[58] but still remains somewhat controversial. The early period has been summarized in the paper by Nozieres and Lewiner,[59] where a set of competing (and partly canceling) terms has been derived and compared for a centrosymmetric semiconductor. They attributed AHE to extrinsic mechanisms. Remarkably, mean free time $\tau$ drops out from the side jump that therefore depends only on the parameters of a perfect crystal; this conclusion agrees with the previous result by Luttinger.[60] More recently, AHE has been related to a Berry phase in $\mathbf{k}$-space that is essentially intrinsic,[61-63] and it seems probable now that Berry curvature is an elegant mathematical language for describing the mechanism that in simplified models was appreciated as side jump. In the framework of Boltzmann equation, side jump appears as the next order correction, in the small parameter $\hbar/E_F\tau$, to the skew scattering term in the Hall conductivity $\sigma_{xy}$; $E_F$ is Fermi energy. In the meantime, some experimental data suggest that this correction term dominates in the dirty regime.[60,64] A topological protection of the side jump contribution to $\sigma_{xy}$ seems to be the most natural explanation of its remarkable ubiquity. However, the fact that side jump contribution has, in the framework of Ref. 59, the same magnitude but opposite sign in the clean and dirty limits, indicates that the problem still persists.

The problems that make theory of AHE so tricky are also inherent in the theory of spin Hall effect. Moreover, while the definition of the anomalous Hall current is straightforward, the ambiguity of the spin current concept makes calculating the spin Hall effect much trickier. It has been shown that the data of Ref. 25 can be reasonably described by the extrinsic mechanism,[65,66] while the data of Ref. 26 seems to indicate the role of intrinsic mechanisms.[67] Remarkably, the side jump term of Ref. 65 coincides with Berry curvature ($\nabla_{\mathbf{k}} \times \mathbf{r}_{so}$), where $\mathbf{r}_{so}$ is the spin-orbit contribution to the operator of coordinate in the crystal-momentum representation. From this standpoint, side jump can be understood as an intrinsic effect that originates from the operator $\mathbf{r}$ rather than the Hamiltonian $H(\mathbf{k})$. Meanwhile, there is no doubt that in noncentrosymmetric crystals $H(\mathbf{k})$ contributes to spin transport, and this contribution cannot be expressed in terms of Berry curvature. Indeed, $\mathbf{r}_{so} = (u_{\mathbf{k}}| i\nabla_{\mathbf{k}} |u_{\mathbf{k}})$ is exactly the same for the Hamiltonians $H_\alpha$ with $\alpha = $ const and $\alpha = \alpha(k^2)$, while spin currents do vanish in the first case and do not in the second;[68] here $u_{\mathbf{k}}$ are eigen-spinors. The same is valid for Hamiltonians with parabolic and nonparabolic $H_0$ parts.[69] Also, $\nabla_{\mathbf{k}} u_{\mathbf{k}}$ is not defined at $\mathbf{k} = 0$ for the Hamiltonians of $H_\alpha$ type.

It has been shown recently,[70] that the joint effect of intrinsic and extrinsic terms in $H(\mathbf{k})$ on spin currents is singular. In $H_\alpha$ semiconductors, spin current $j_x^z$ defined according Eq. (10) vanishes for arbitrary $\alpha \neq 0$, i.e., spin precession in the field $\mathbf{B}_\alpha$ nullifies the extrinsic spin current after the integration over the whole specimen. This can be understood as the result of averaging the spins, polarized by skew scattering, over the electron trajectories, and seems to underscore the fact that spin accumulation near boundaries cannot be derived from spin currents of Eq.

(10). The same conclusion comes from the observation that spin relaxation on the boundary supports spin Hall effect even when bulk spin currents vanish.[71,72]

Analysis of the existing data on spin currents and spin Hall effect suggest that, at a qualitative level, they can be better related if, instead of $q = 0$ components of spin currents corresponding to averaging over the entire infinite homogeneous space, one considers their Fourier components at the momenta $q \approx k_\alpha$. Such an approach corresponds to the idea that when it comes to spin accumulation $S$ at the edge, only the adjacent layer of the width about $\ell_\alpha$ matters. Fourier components $j_i^z(k_\alpha)$ do not vanish for the $H_\alpha$ Hamiltonian and have the same magnitude of about $eE/\hbar$ as for the Hamiltonians that are nonlinear in $k$ (e.g., the $k^3$ heavy hole spin-orbit Hamiltonian),[73] if to generalize the definition of $k_\alpha$ by expressing it in terms of the spin-orbit splitting $\delta_{so}$ at the Fermi level, $k_\alpha \to k_{so} = m\delta_{so}/2\hbar^2 k_F$. Moreover, spin Hall currents $j_{sH}$ defined in such a way can be related to spin accumulations as

$$S/\hbar \sim k_{so}\tau\, j_{sH}(k_{so}), \quad j_{sH}(k_{so}) \sim eE/4\pi\hbar. \tag{11}$$

With such redefinition of spin currents, they acquire some universality in establishing the basic scales and connection to spin accumulations near the edges.[72] Equation (11) shows that $j_{sH}(k_{so})$ coincides by the order of magnitude with the spin current by Sinova et al.[46] but has somewhat different physical meaning. Numerical constants in Eq. (11) essentially depend on the specific form of spin-orbit coupling and on boundary conditions and can be only found from detailed transport equations. There is no doubt that physical quantities like $S$ are continuous functions of all parameters, including $\alpha$. Also, near the sample edge, spin magnetization $S(x)$ is an oscillating function of $x$ (with a period about $k_{so}^{-1}$) that usually changes sign, $x$ being separation from the edge. Hence, it is difficult to expect existence of any universal relation even between the signs of the bulk spin current and the spin accumulation near the edge.

## 7. Conclusions

Spin-orbit coupling is currently considered as a key for creating and manipulating spin populations electrically, at a nanometer scale. Recent years have witnessed impressing progress in this field, both in experiment and in theory. The very possibility of creating nonequilibrium spins by electric fields is based on spin nonconservation. This fact, in turn, results in an essential difference between the spin-transport theory in media with spin-orbit coupling and the traditional theory of charge transport. Recent progress in the theory and response to emerging challenges are discussed.

**Acknowledgments**

This work was supported by Harvard Center for Nanoscale Systems. Inspiring discussions with H.-A. Engel, B. I. Halperin, A. H. MacDonald, C. M. Marcus, D. Loss, and Q. Niu are gratefully acknowledged.